\documentclass[aps,prd,preprint,groupedaddress]{article}
\pdfoutput=1
\usepackage{jcappub}
\usepackage{multirow, amsmath, amssymb, color, array}

\begin{document}

\title{Sensitivity of IceCube-DeepCore to neutralino dark matter in the MSSM-25}

\author[a]{Hamish Silverwood,}
\author[b]{Pat Scott,}
\author[c]{Matthias Danninger,}
\author[c,d]{Christopher Savage,}

\author[c]{Joakim Edsj\"{o},}

\author[a]{Jenni Adams,}
\author[a]{Anthony M Brown}
\author[c]{and Klas Hultqvist}

\affiliation[a]{Department of Physics and Astronomy, University of Canterbury, Christchurch 8140, New Zealand}
\affiliation[b]{Department of Physics, McGill University, Montr\'eal QC H2W2L8, Canada}
\affiliation[c]{Oskar Klein Centre for Cosmoparticle Physics, Department of Physics, Stockholm University, SE-10691 Stockholm, Sweden}
\affiliation[d]{Department of Physics \& Astronomy, University of Utah, Salt Lake City, UT 84112, USA}

\date{\today}

\begin{abstract}
{We analyse the sensitivity of IceCube-DeepCore to annihilation of neutralino dark matter in the solar core, generated within a 25 parameter version of the minimally supersymmetric standard model (MSSM-25). We explore the 25-dimensional parameter space using scanning methods based on importance sampling and using \texttt{DarkSUSY 5.0.6} to calculate observables. Our scans produced a database of 6.02 million parameter space points with neutralino dark matter consistent with the relic density implied by WMAP 7-year data, as well as with accelerator searches. We performed a model exclusion analysis upon these points using the expected capabilities of the IceCube-DeepCore Neutrino Telescope. We show that IceCube-DeepCore will be sensitive to a number of models that are not accessible to direct detection experiments such as SIMPLE, COUPP and XENON100, indirect detection using Fermi-LAT observations of dwarf spheroidal galaxies, nor to current LHC searches.} 

\end{abstract}

\maketitle

\section{Introduction}
Much of the development of supersymmetry (SUSY) was motivated by problems within the Standard Model (SM), such as the fine tuning of the Higgs mass \citep{Witten1981, Dimopoulos1981} and the unification of gauge coupling constants at high energy \citep{Ellis1991, Amaldi1991, Langacker1991, Giunti1991, Ross1992}. When the lightest neutralino is the lightest supersymmetric particle (LSP) and $R$-parity is conserved, the lightest neutralino is a natural candidate for dark matter, providing a solution to one of the most important problems in astrophysics. 

Neutralinos can accumulate in the centre of astrophysical bodies such as the Sun, where they would annihilate into a range of SM particles. Subsequent decays and interactions of these SM particles would yield neutrinos, which have the unique ability to escape the solar core and reach detectors on Earth \citep{Press1985, Silk1985, Krauss1985}. 

One such detector is the IceCube Neutrino Telescope \citep{Halzen2010}. Previous analyses of data recorded by IceCube in the 22- and 40-string configurations have already provided limits on neutralino dark matter annihilation in the Sun, the neutralino-proton cross-section, and the resulting muon flux \citep{IC22muonflux2009, Abbasi2012}. Further papers have used projected capabilities of the full 86-string IceCube-DeepCore detector (IceCube-86) to analyse parameterisations of the MSSM in more detail, such as the four-parameter Constrained MSSM (CMSSM) \citep{Trotta2009, Ellis2010}, the seven parameter MSSM-7 \citep{Wikstrom2009, IC22methods}, and the 19 parameter phenomenological MSSM (pMSSM) \citep{Cotta2011}.

In this paper we present the results of an exploration of a 25 parameter version of the MSSM and the sensitivity of IceCube-86 to these models. This analysis involves two new elements compared to previous work: the parameter space has been enlarged to 25 parameters, reducing the artificial constrictions placed upon the MSSM; and the IceCube-86 model exclusion confidence level (CL) calculations now include point-by-point optimisation of the reconstructed cut angle around the Sun, which was not included in a similar analysis performed in \citep{IC22methods}.  

The structure of this paper is as follows. In Section \ref{SectionIceCube} we introduce the IceCube Neutrino Telescope and its salient characteristics. Then, in Section \ref{SectionMethods}, we present the details of our analytical methods: the MSSM-25 parameterisation within which we worked (\ref{SubsectionMSSM25Parameterization}), the method by which we explored this parameter space (\ref{SubsectionDarkSUSYCalculations}), the calculation of signal rates in IceCube (\ref{SubsectionIceCubeSignalRates}), and finally the calculation of IceCube model exclusion CLs (\ref{SubsectionLikelihoodCalculation}). In Section \ref{SectionResultsDiscussion} the results of this exploration and analysis are presented and discussed.

\section{The IceCube Neutrino Telescope} 
\label{SectionIceCube}
Completed in December 2010, the IceCube Neutrino Observatory consists of a cubic kilometre of extremely transparent natural ice deep below the South Pole, instrumented with 5160 digital optical modules (DOMs), along with 320 DOMs embedded in ice-filled tanks at the surface (known as the IceTop array). The in-ice DOMs are arranged along 86 strings lowered down vertical wells melted into the ice using a hot water drill. Seventy-eight of the strings are arranged in a hexagonal grid with a horizontal spacing of 125\,m between strings, with DOMs  separated vertically by 17\,m and situated between 1450\,m and 2450\,m below the surface. The remaining eight strings are clustered in the centre with a horizontal spacing of about 72\,m or less, and have 50 of their DOMs positioned with 7\,m vertical spacing starting at the bottom. The remaining ten are arranged higher up the strings to act as a down-going muon veto. These eight strings and the 12 adjacent standard IceCube strings make up the DeepCore subarray. DeepCore provides increased sensitivity at lower energies and reduces the energy threshold to approximately 10\,GeV \citep{AbbasiDeepCore2012}. The IceTop array is used to study cosmic ray flux and composition, and serves as a veto for high energy cosmic ray showers \citep{Halzen2010}.

IceCube can reconstruct the type, energy and direction of the incoming neutrinos by observing the \v{C}erenkov radiation emitted by leptons created in charged current interactions between the neutrinos and nuclei in the ice. Muons produced in such interactions are of particular interest, as their mean free paths can be up to 10km for the most energetic neutrinos. One can thus detect neutrinos that interact outside the volume of instrumented ice making up IceCube \citep{Halzen2010}. 

The sensitivity of IceCube to neutrinos of different energies is characterised by the effective area. For this study we use the effective area for muon neutrino detection by the 86-string IceCube configuration (IceCube-86), derived from simulation and presented in \citep{Danninger2011}. The sensitivity study presented in~\citep{Danninger2011} is for the full 86 string detector with the initially proposed DeepCore geometry, that consisted of only six additional strings instead of the eight deployed. The study was performed as a full analysis in all details on simulated backgrounds, including realistic data processing and event selection. After the work presented in this paper was completed, preliminary results from the IceCube-79 detector were released \citep{IDM_talk_Matthias}; final results were submitted just before this paper was accepted \citep{IceCubeSolarDarkMatter2012}. A comparison of the effective area from \citep{Danninger2011} and these more recent results suggests a moderate over-estimation of the effective area by the earlier sensitivity study: a factor of $\sim$3 at 100\,GeV neutrino energy, dropping to $\sim$1.5 at 1\,TeV. Given the order unity difference, the effective area from the IceCube-86 sensitivity analysis \citep{Danninger2011} will suffice for our purposes here. 

\section{Methods} \label{SectionMethods}

\subsection{The MSSM-25 Parameterization} \label{SubsectionMSSM25Parameterization}

The full MSSM has 178 free parameters: 158 from the soft SUSY-breaking sector, one from the coupling of the two Higgs supermultiplets, and the remaining 19 from the Standard Model \citep{BaerTata2006}. 54 of these parameters can be removed by suppressing the $\mathbf{C}$-terms of the soft SUSY-breaking Lagrangian, leaving 124 parameters. We direct the reader to \citep{BaerTata2006} for more information. For phenomenological studies it is useful to reduce this number in order to make exploration of the parameter space computationally feasible. This reduction is performed via a series of assumptions that become progressively more severe as more parameters are eliminated. Previous analyses of neutralino dark matter signals in IceCube have used models with four, seven, nine and 19 parameters \citep{Wikstrom2009, Trotta2009, Ellis2010, Cotta2011, IC22methods}. For this research we used a 25-parameter version of the MSSM.

The potential utility of increasing the number of parameters lies in reducing the artificial constrictions placed upon the MSSM and allowing different combinations of parameters that could perhaps yield dark matter candidates with previously unseen properties. The disadvantage of increasing the number of parameters is the increased computation time needed to sample the parameter space.

The parameters of the MSSM-25 are listed in the left column of Table \ref{MSSM25SimulationParameters}. The Higgs sector is described by the ratio of Higgs vacuum expectation values (VEVs) $\tan{\beta}$, the CP-odd Higgs mass $m_A$ generated after electroweak symmetry breaking, and the $\mu$ parameter from the MSSM superpotential. The gaugino sector of the soft SUSY-breaking Lagrangian $\mathcal{L}_\text{soft}$ is parameterised by the three mass terms $M_1$, $M_2$, and $M_3$.

The five Hermitian $3\times3$ sfermion mass-squared matrices in $\mathcal{L}_\text{soft}$ are simplified by assuming the absence of flavour changing neutral currents (FCNCs), forcing all non-diagonal entries to zero. Each mass squared matrix therefore only has three real free parameters, and so the sfermion sector has a total of 15 parameters. 

For the triple scalar couplings in $\mathcal{L}_\text{soft}$ we again apply the prohibition on FCNCs to eliminate the off-diagonal entries. We also restrict the matrices to be Hermitian (and so the diagonal elements to be real) in order to eliminate any CP-violating phases \citep{Chung2005}. We also make the assumption that the elements of the triple scalar coupling matrices are proportional to the corresponding elements of the Yukawa matrices $\mathbf{y}_\text{u}$, $\mathbf{y}_\text{d}$, and $\mathbf{y}_\text{e}$. We can set the first and second entries of $\mathbf{a}_\text{u}$ and $\mathbf{a}_\text{d}$ to zero as the corresponding Yukawa couplings are negligibly small, but we retain one parameter, $a_{\text{e}1/2}$, for the first and second entries of $\mathbf{a}_\text{e}$, as it is relevant to the calculation of the anomalous magnetic moment of the muon \citep{Abdussalam2010}. We finally arrive at the following triple scalar coupling matrices:

\begin{align}
\mathbf{a}_\text{u} &= \left( \begin{array}{ccc}
\phantom{ccc}0\phantom{ccc} & 0  &0   \\
 0 & 0 &0   \\
 0 & 0  & a_{\text{u}3} Y_{u33}  
\end{array} \right)\hspace{0.5cm}
\mathbf{a}_\text{d} &= \left(\begin{array}{ccc}
 \phantom{ccc}0\phantom{ccc} & 0  &0   \\
 0 & 0 &0   \\
 0 & 0  & a_{\text{d}3} Y_{d33}
\end{array}\right) \hspace{0.5cm}
\mathbf{a}_\text{e} &= \left( \begin{array}{ccc}
 a_{\text{e}1/2} Y_{e11} & 0  &0   \\
 0 & a_{\text{e}1/2} Y_{e22}  &0   \\
 0 & 0  & a_{\text{e3}} Y_{e33}  
\end{array}\right)
\end{align}

\subsection{Parameter Scans} \label{SubsectionDarkSUSYCalculations}

We performed scans of this parameter space using importance sampling with ADSCAN \citep{Brein2004}, an adaptive scanning programme based on the VEGAS algorithm \citep{Lepage1978}.  The VEGAS algorithm works by performing an initial random scan of the parameter space and then focusing subsequent scans on areas that produce important values, as defined by an importance function $G$. Our importance function was 
\begin{align}
G\left(\Omega_\chi h^2\right) = \text{exp}\left[-\frac{1}{2} \left(\frac{\Omega_\chi h^2 - \Omega_{\text{WMAP-7}} h^2}{\sigma_{\Omega h^2}}\right)^2 \right]
\end{align}
where $\Omega_{\text{WMAP-7}} h^2 = 0.1120 $, the cold dark matter relic density derived from 7-year WMAP measurements, with $1\sigma$ uncertainty of $\pm 0.0056$ \citep{Larson2011}. For the error in the denominator we chose $\sigma_{\Omega h^2} = 0.01$ to allow for some theoretical error above the $1\sigma$ experimental error of $\pm 0.0056$, as we did in the corresponding MSSM-7 ADSCAN parameter scans in \citep{IC22methods}. The parameter $\sigma_{\Omega h^2}$ determines the width of the Gaussian importance function used in the scanning process, with larger numbers generating a broader scan. 

We ran the adaptive scanning programme multiple times with varying parameter limits, which are listed in Table \ref{MSSM25SimulationParameters}. We rejected points in the parameter space that gave negative mass-squared terms, and where the lightest neutralino was not the LSP. We eliminated additional points by applying experimental limits on the accelerator observables produced by each point: the sparticle mass spectrum \citep{PDG2002}; standard model and SUSY Higgs masses from LEP and CDF measurements \citep{PDG2002}; the $b \rightarrow s\gamma$ branching ratio using results from the Heavy Flavour Averaging Group \citep{2006HFAG}; the invisible $Z^0$ boson decay width \citep{PDG2002}; and the magnetic moment of the muon derived from data from the BaBar and KLOE experiments \citep{Davier2009, Davier2010, Teubner2010}. These limits are the defaults used by DarkSUSY. 

As the primary focus of this paper is the impact of IceCube-DeepCore, our overall strategy was to deliberately lean towards a more inclusive scan, choosing to discard only points that disagreed with the most established and well-understood experimental limits. We made this choice so as to keep scans simple and easy to understand, as this helps to more clearly focus on the specific impacts of IceCube-DeepCore on the MSSM-25.  In this way we avoid worrying about erroneously discarding some models due to e.g.\ the unknown model-dependence of existing LHC SUSY limits when translated into the MSSM-25, or the massive amount of computational resources required to obtain this knowledge.

From the database of points satisfying physicality, the demand for a neutralino LSP, and accelerator bounds, we selected those that gave a relic density within the $2\sigma$ error bounds of the 7-year WMAP measurement, i.e.\ in the range $0.1120 - 0.0112 \le \Omega_\chi h^2 \le 0.1120 +0.0112$. We found a total of 6.02 million such points within our original database. \footnote{These points are available online at \url{http://grappa.science.uva.nl/silverwood/MSSM25_Database/HS_MSSM25_2013_Database.zip}.} 

For calculating the values of all observables, we used the \texttt{DarkSUSY 5.0.6} software package \cite{DarkSUSY2009, IC22methods}, with default options except where otherwise specified.  In particular, for the calculation of neutralino-nucleon scattering cross-sections this implies $\Delta_s=-0.12$ for the strange quark content of the nucleus, and $\Sigma_{\pi N} =45$\,MeV, $\sigma_0 =35$\,MeV for the hadronic matrix elements.

\begin{table}
\begin{center}
\begin{tabular}{l c c c }

\hline
Parameter & Units &\multicolumn{2}{c}{Limits}\\
\hline
Gaugino mass terms $M_1$, $M_2$, and $M_3$ & GeV & $-$16000  & 16000 \\
& & & \\
Higgs parameter $\mu$ & GeV & $-$100000  & 100000 \\
CP-odd Higgs mass $m_A$ & GeV & $-$8000  & 8000 \\
Higgs VEV ratio $\tan{\beta}$ & & 2  & 65 \\
& & & \\
Sfermion mass parameters $m_{\tilde{Q}_{1}}$, $m_{\tilde{Q}_{2}}$,  & \multirow{3}{*}{GeV} & \multirow{3}{*}{100} & \multirow{3}{*}{25000}\\
$m_{\tilde{Q}_{3}}$, $m_{\tilde{\bar{u}}_1}$, $m_{\tilde{\bar{u}}_2}$, $m_{\tilde{\bar{u}}_3}$, $m_{\tilde{\bar{d}}_1}$, $m_{\tilde{\bar{d}}_2}$, $m_{\tilde{\bar{d}}_3}$, \\
$m_{\tilde{L}_{1}}$, $m_{\tilde{L}_{2}}$, $m_{\tilde{L}_{3}}$, $m_{\tilde{\bar{e}}_1}$, $m_{\tilde{\bar{e}}_2}$ , and $m_{\tilde{\bar{e}}_3}$  & & & \\
& & & \\
Triple scalar couplings $a_{\text{u}3}$, $a_{\text{d}3}$, $a_{\text{e}1/2}$, and $a_{\text{e3}}$ & GeV & $-$50000  & 50000 \\

\hline
\end{tabular}
\caption{MSSM-25 Parameter Space Maximum Exploration Limits}
\label{MSSM25SimulationParameters}
\end{center}
\end{table}

\subsection{IceCube Signal Rates} 
\label{SubsectionIceCubeSignalRates}

We post-processed the 6.02 million points with suitable relic density using \texttt{DarkSUSY 5.0.6}, as described in \cite{IC22methods}, to find the IceCube signal rates that would be produced by each model. 
 
We start by modelling the accumulation and annihilation of neutralinos in the Sun. The total population of neutralinos in the Sun $N(t)$ is described by the equation \citep{Jungman1996} 
\begin{align}
\frac{\text{d} N(t)}{\text{d}t} = C_c - C_a N(t)^2  \label{WIMPdiffeq}.
\end{align}
where $C_c$ is the constant neutralino capture rate and $C_a$ is defined as
\begin{align}
C_a = \langle \sigma_a v \rangle \frac{V_2}{V_1^2} \hspace{0.5cm} \text{with} \hspace{0.5cm}
V_j \approx 6.5 \times 10^{28} \left( \frac{j m_\chi}{10 \text{GeV}} \right) ^{-3/2} \text{cm}^3.
\end{align}
The $V_j$ terms are effective volumes, which take into account the quasi-thermal distribution of neutralinos in the Sun, and $\langle \sigma_a v \rangle$ is the total self annihilation cross section multiplied by the relative velocity, taken in the limit of zero relative velocity. 

The capture rate calculation depends on the halo density and velocity profile of dark matter, the neutralino mass and interaction cross-section, and the properties of the Sun; see e.g.\ \citep{Scott2009} for a detailed description. We assume a standard dark matter halo model, with the Sun moving at $v_\odot = 220$\,km\,s$^{-1}$ through a halo with local dark matter density $\rho_0 = 0.3$\,GeV\,cm$^{-3}$, and dark matter velocities following a Maxwell-Boltzmann distribution with dispersion $\bar{v}=270$\,km\,s$^{-1}$.  We did not include the detailed effects of diffusion and planets upon the capture rate, as the simple free-space approximation \cite{Gould1987} included in DarkSUSY turns out to be highly accurate \citep{Sivertsson2012}.  The annihilation rate of neutralino pairs is given by 
\begin{align}
\Gamma_a (t) = \frac{1}{2} C_a N^2(t).
\end{align}
Solving Equation \ref{WIMPdiffeq} gives us the annihilation rate at a given time $t$ as 
\begin{align}
\Gamma_a (t) = \frac{C_c}{2} \tanh^2 \left(\frac{t}{\tau} \right) \label{AnnihilationRateEq}
\end{align}
where
\begin{align}
\tau = \frac{1}{\sqrt{C_c C_a}}
\end{align}
is the capture-annihilation equilibrium time scale.  Taking $t$ as the age of the solar system (4.5 billion years) and approximating the structure of the Sun to be constant over its lifetime gives the current neutralino annihilation rate in the Sun.

From the annihilation rate $\Gamma_a(t)$ and branching fractions, \texttt{DarkSUSY} calculates a predicted neutrino flux spectrum at the IceCube detector, using lookup tables previously computed with WimpSim \cite{Blennow2008}.  WimpSim takes neutrino production in the Sun and propagation through the Sun, Earth and interplanetary space into account using a full three-flavour Monte Carlo simulation. Together with the effective area of IceCube-86 and the mean angular error obtained from \citep{Danninger2011}, and some choice of analysis region around the Sun, the calculated flux yields an expected signal rate for any given model.

The other necessary ingredient for our model exclusion calculations is the number of background events. We generated these using a bootstrap Monte-Carlo re-simulation of the expected background rates away from the Sun, as derived in the sensitivity study~\citep{Danninger2011}. The total expected number of background events over the whole sky was 15552 \citep{DanningerThesis2011}, consisting of atmospheric neutrino and muon backgrounds. Given only the total number of background events, we assumed all events to be uniformly distributed in declination. This is a good approximation for atmospheric neutrinos within the narrow declination band of the analysis, and a conservative choice for the down-going atmospheric muon component. This results in an overall robust background expectation. The single realisation of this expected background that we employed for this study contained 1452 simulated events within $20^\circ$ of the solar centre.  

\subsection{Likelihood Calculation} \label{SubsectionLikelihoodCalculation}

In this work we perform a model exclusion analysis similar to that of Section 4.3 in \citep{IC22methods}. Compared to \citep{IC22methods}, we employ an expanded parameter space and optimise our angular event cut around the solar position. We also look in far more detail than \citep{IC22methods} into the impacts of IceCube-86 on the model parameters and their corresponding derived observables. To quantify the exclusion CL we calculated the $p$-value for each model using the method outlined in Section 3.6 of \citep{IC22methods}, specifically Equation 3.28 and its contributors. The CL at which a model can be excluded is then $1-p$. 

We excluded all events with reconstructed angle outside some bin of width $\phi$ centred on the Sun. We optimised the cut angle by repeating the $p$-value calculation for $\phi = 3^{\circ}, 6^{\circ}, 9^{\circ}$, and $20^{\circ}$, and selecting the lowest $p$-value (i.e.\ highest exclusion CL) for each model. We optimised the cut angle because different models produce signals with different angular distributions. For instance, higher mass neutralinos tend to produce higher-energy neutrinos, which are more accurately reconstructed by IceCube and are therefore more densely clustered around the solar position.  Using a small cut angle in this case reduces the number of background events included in the bin more than it reduces the number of signal events. A brief analysis of optimal cut angle is presented in Section \ref{SectionResultsDiscussion}.

Because we work entirely with simulated data in this paper, we can simply optimise the cut angle by choosing the value that produces the smallest $p$-value.  Working with real data however, selecting the cut angle that gave the lowest $p$-value would constitute an \textit{a posteriori} analysis choice, which would make the statistical treatment less than rigorous.  As we show later, the cut angle giving the lowest $p$-value is strongly correlated with the neutralino mass and somewhat dependent on the neutralino character, the quantities that define the expected neutrino energy spectrum (this is to be expected, because the angular distribution of muons is heavily dependent upon the neutrino energies).  In an analysis of real data, one should therefore optimise the cut angle in advance using the expected neutrino energy spectrum for a given model, and an ensemble of background-only simulations like the one we employ here.

\section{Results and Discussion} \label{SectionResultsDiscussion}

\begin{figure}[tbp]
	\centering
	\includegraphics[width=\linewidth, trim = 60 0 95 0, clip=true]{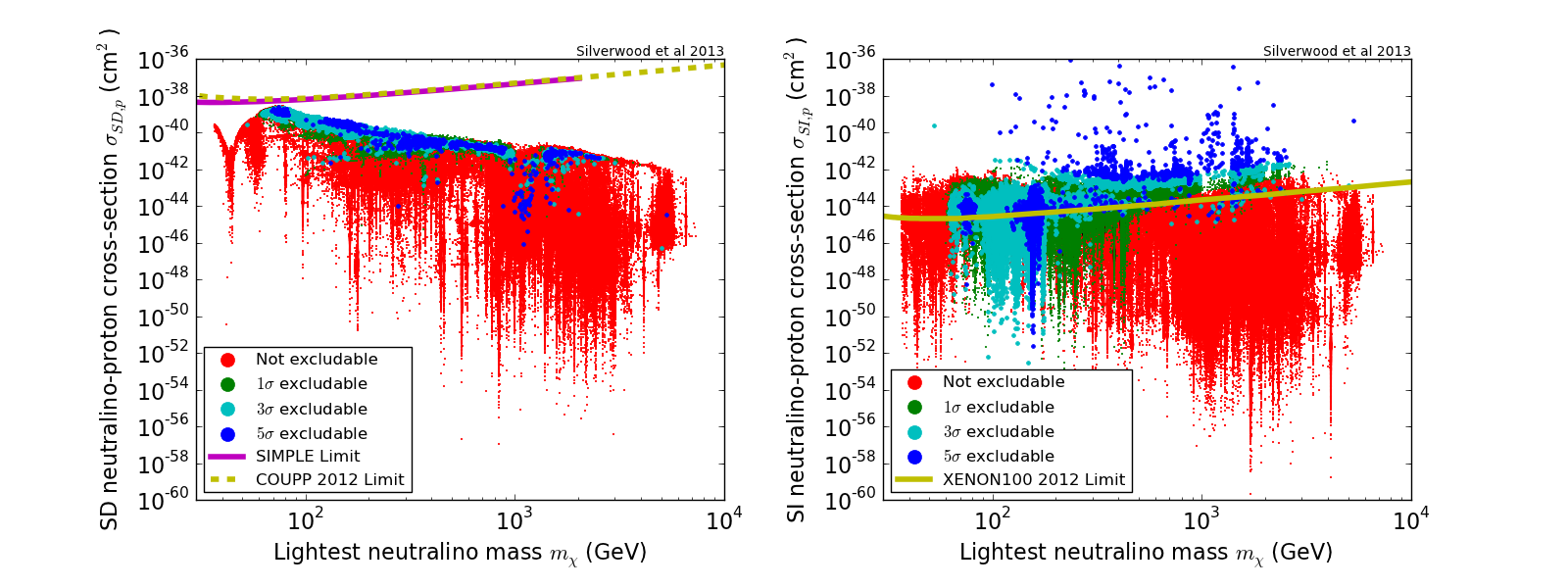}
	\caption{Spin-dependent (SD) neutralino-proton cross-section $\sigma_{\text{SD},p}$ (left), and spin-independent (SI) neutralino-proton cross-section $\sigma_{\text{SI},p}$ (right) as functions of lightest neutralino mass $m_\chi$, for points derived from explorations of the MSSM-25 parameter space. In the left panel 90\% CL spin-dependent WIMP-proton cross-section limits from SIMPLE \citep{Felizardo2011} and COUPP \citep{COUPP2012} direct detection experiment are displayed as magenta and yellow lines respectively. In the right panel the 90\% CL spin-independent WIMP-nucleon cross-section limit from 225 live days of XENON100 direct detection experiment data is displayed as a yellow line  \citep{Aprile2012}. Above $m_\chi$ = 1\,TeV the XENON100 limit is based on points from the XENON100 collaboration. Colour coding indicates predicted IceCube-86 model exclusion CL. The areas of cyan and blue points show that IceCube-86 has the ability to exclude models beyond the reach of current direct detection experiments such as SIMPLE, COUPP, and XENON100.}
	\label{NMvDD8_Neutralino_mass_v_SDp_SIp}
\end{figure}

\begin{figure}[tbp]
	\centering
	\includegraphics[width=\linewidth, trim = 60 0 95 0, clip=true]{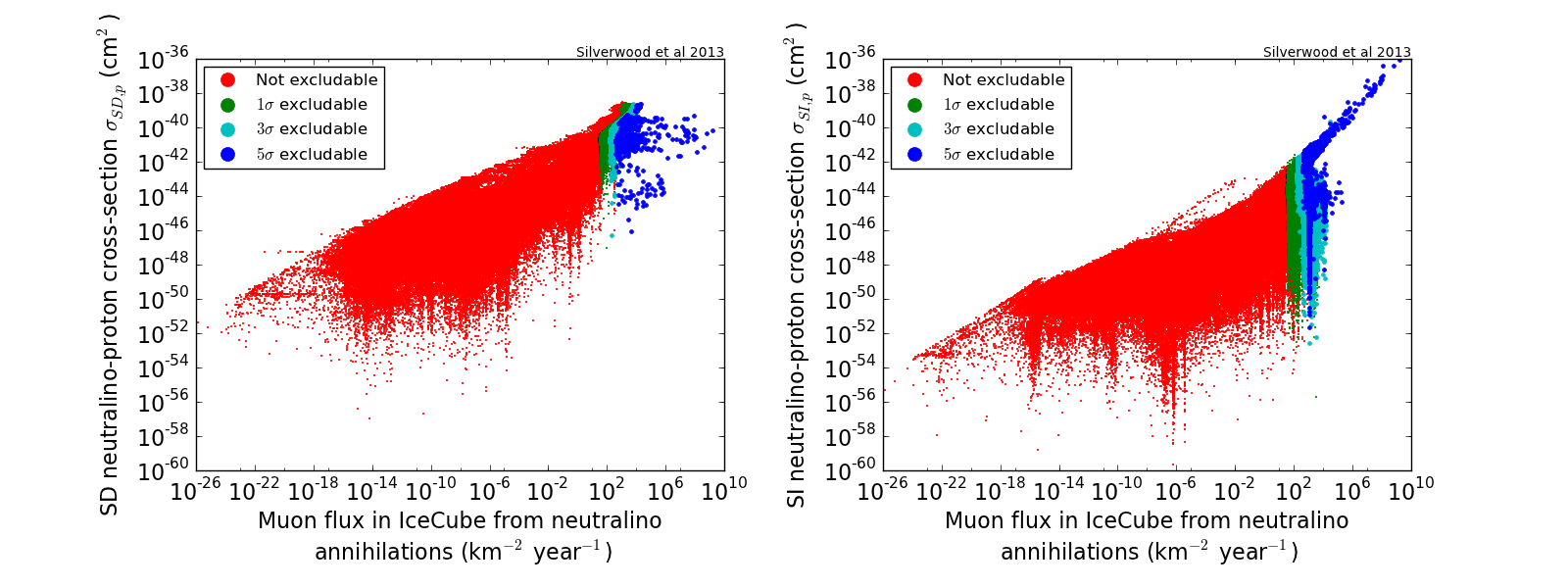}
	\caption{Spin-dependent neutralino-proton cross-section $\sigma_{\text{SD},p}$ (left) and spin-independent neutralino-proton cross-section $\sigma_{\text{SI},p}$ (right) against muon flux in IceCube from neutralino annihilations, for points derived from explorations of the MSSM-25 parameter space. Colour coding indicates predicted IceCube-86 model exclusion CL.}
	\label{NMvDD7_Solar_Muon_Rate_v_SDp_SIp}
\end{figure}

Figure \ref{NMvDD8_Neutralino_mass_v_SDp_SIp} shows the results of the model exclusion analysis we performed on our set of 6.02 million points. Care must be taken in interpreting these plots for two reasons. First, we plot points with exclusion CL of $3\sigma$ and above (cyan and blue) with larger dots than points with lower exclusion CL, as points with higher exclusion CL are less numerous but also more important to our discussions, and so need to be emphasized. Second, we plot points of higher exclusion CL on top of points with lower exclusion CL, and so the former can obscure the latter. An area of cyan or blue on Figure \ref{NMvDD8_Neutralino_mass_v_SDp_SIp} means that IceCube has exclusion (or detection) capability of $3\sigma$ (99.7\% CL) or better, for a certain range of interaction cross-sections, neutralino masses, \textit{and} other MSSM-25 parameters. Nevertheless, one can see in the areas of blue and cyan points that IceCube has the ability to exclude models at better than 99\% CL well beyond even the 90\% CL reach of current direct detection experiments such as SIMPLE \citep{Felizardo2011}, COUPP \citep{COUPP2012} and XENON100 \citep{Aprile2011, Aprile2012}.

In Figure \ref{NMvDD7_Solar_Muon_Rate_v_SDp_SIp} we see spin-dependent (SD) and spin-independent (SI) neutralino-proton cross-sections plotted against the predicted muon flux in IceCube-86 from solar neutralino annihilations.  For these calculations we used a muon energy threshold of 1 GeV, and a maximum angular separation of 30$^\circ$ between the solar centre and muon arrival angle.

\begin{figure}[tbp]
	\begin{minipage}[t]{0.5\linewidth}
		\centering
		\includegraphics[width=\textwidth, trim = 0 10 40 0, clip=true]{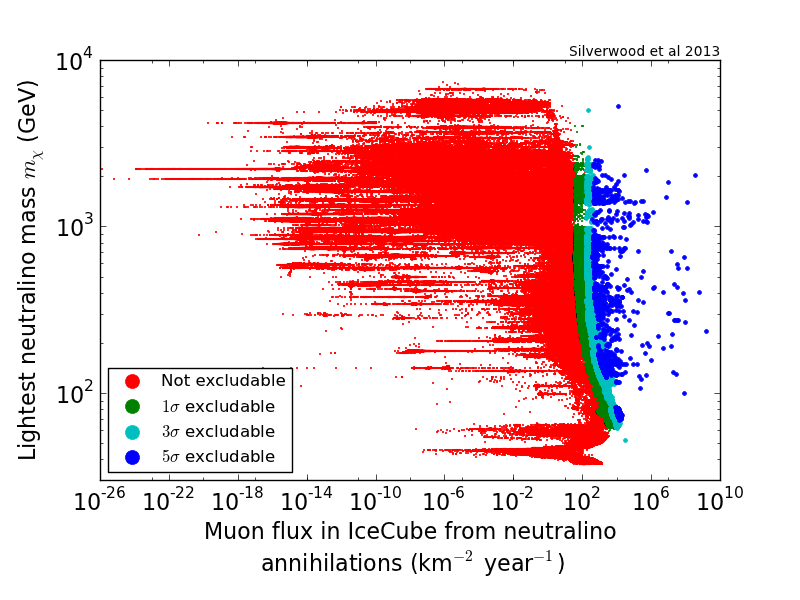}
		\caption{Lightest neutralino mass $m_\chi$ against muon flux in IceCube from solar neutralino annihilations, for points derived from explorations of the MSSM-25 parameter space. Colour coding indicates predicted IceCube-86 model exclusion CL. The effect of the energy dependence of IceCube's effective area can be seen in the distribution of high exclusion CL points: lower-mass neutralinos will yield lower energy neutrinos/muons, and so a higher flux is necessary to create the same signal and thus exclusion CL in IceCube. }
		\label{NMvDD9_IceCube_Muons_v_Neutralino_mass}
	\end{minipage}
	\hspace{0.5cm}
	\begin{minipage}[t]{0.5\linewidth}
		\centering
		\includegraphics[width=\textwidth, trim = 0 10 40 0, clip=true]{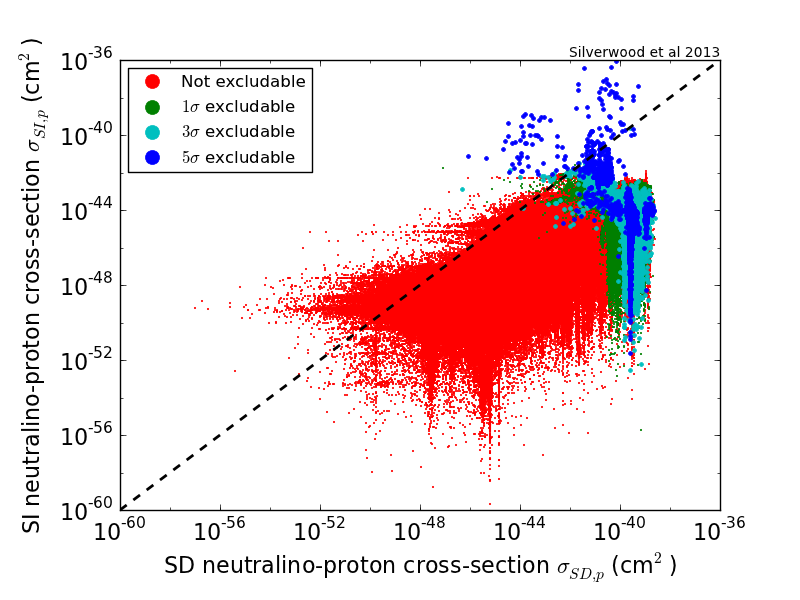}
		\caption{Spin-independent neutralino-proton cross-section $\sigma_{\text{SI},p}$ against spin-dependent neutralino-proton cross-section $\sigma_{\text{SD},p}$ for points derived from explorations of the MSSM-25 parameter space. Colour coding indicates predicted IceCube-86 model exclusion CL. Dashed black line indicates equality between spin-independent and spin-dependent neutralino-proton cross-sections.}
		\label{NMvDD5_SD_proton_Xsec_v_SI_proton_Xsec}
	\end{minipage}
\end{figure}

In general models which can be excluded at higher CL have higher muon fluxes -- high muon flux leads to high signal, and high signal models are easier to rule out. This relationship, however, is not exact, as witnessed by the fact that the different regions of exclusion CL (i.e.\ different coloured points) are not cleanly banded. In Figure \ref{NMvDD7_Solar_Muon_Rate_v_SDp_SIp} one can see points with lower exclusion CL that nonetheless have higher muon fluxes than points with higher exclusion CL. This is a result of IceCube's effective area increasing with energy, as seen in Figure 2 of \citep{Danninger2011}. A model can have a high muon flux but if it also has a comparatively low neutralino mass, then it will produce lower energy neutrinos and muons, which leads to a lower IceCube signal rate. Figure \ref{NMvDD9_IceCube_Muons_v_Neutralino_mass} throws this interplay into stark relief; models with a lower-mass neutralino will produce neutrinos with lower energy, and so compared to models with a higher neutralino mass a higher flux is necessary to compensate for the reduced effective area at lower energies. 

Figure \ref{NMvDD5_SD_proton_Xsec_v_SI_proton_Xsec} compares spin-independent to spin-dependent neutralino-proton cross-sections. Note that many models have spin-dependent neutralino-proton cross-sections that are larger than their spin-independent neutralino-proton cross-sections. The majority of models with high exclusion CL have high spin-dependent cross-sections. The long vertical band of high exclusion CL at high spin-dependent cross-section are regions where spin-dependent scattering is the dominant capture mode of neutralinos in the sun. As one moves upwards along this band a `turn-off' is reached at approximately $\sigma_{SI,p} = 10^{-44}$\,cm$^2$; above this point, spin-independent nuclear scattering is a significant source of neutralino capture in the Sun, and even dominates the capture rates in many cases.

\begin{figure}
	\centering
	\includegraphics[width=\linewidth, trim = 60 10 95 0, clip=true]{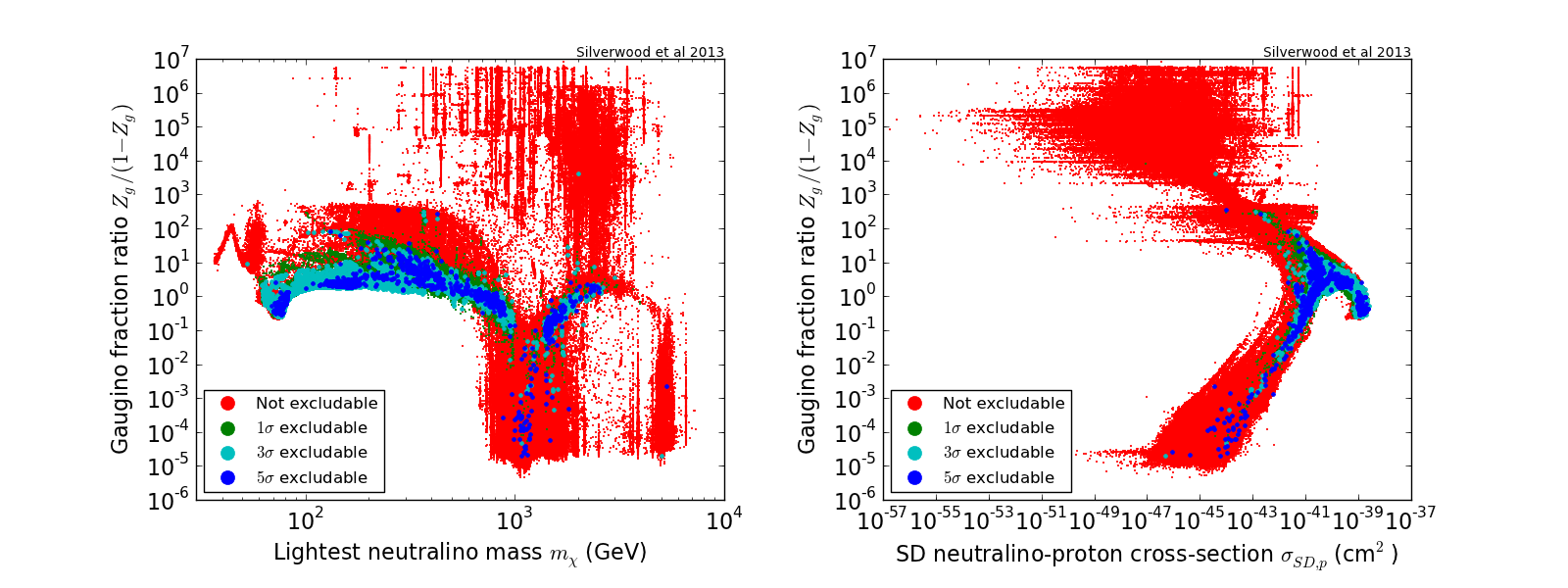}
	\caption{Gaugino fraction ratio $Z_g / (1-Z_g)$ of the lightest neutralino against lightest neutralino mass $m_\chi$ (left) and spin-dependent neutralino-proton cross-section $\sigma_{\text{SD},p}$ (right), for points derived from explorations of the MSSM-25 parameter space. Colour coding indicates predicted IceCube-86 model exclusion CL. Note that the largest spin-dependent cross-sections, and therefore the best IceCube exclusion CL, occur where the lightest neutralino is a mixed gaugino-Higgsino.}
	\label{ArticlePlots2_Gaugino_Fraction_v_Neutralino_mass_SDp}
\end{figure}

As shown in the left panel of Figure \ref{ArticlePlots2_Gaugino_Fraction_v_Neutralino_mass_SDp}, IceCube has its strongest exclusion capability in the region where the neutralino is approximately equal parts gaugino and Higgsino. This corresponds to high spin-dependent neutralino-proton cross-section, as seen in the right panel of Figure \ref{ArticlePlots2_Gaugino_Fraction_v_Neutralino_mass_SDp}. This agrees with the findings found in Section 4.3 of \cite{IC22methods}, where a similar analysis was performed using MSSM-7 models. 

\begin{figure}[tbp]
	\begin{minipage}[t]{0.5\linewidth}
		\centering
		\includegraphics[width=\textwidth, trim = 0 10 40 0, clip=true]{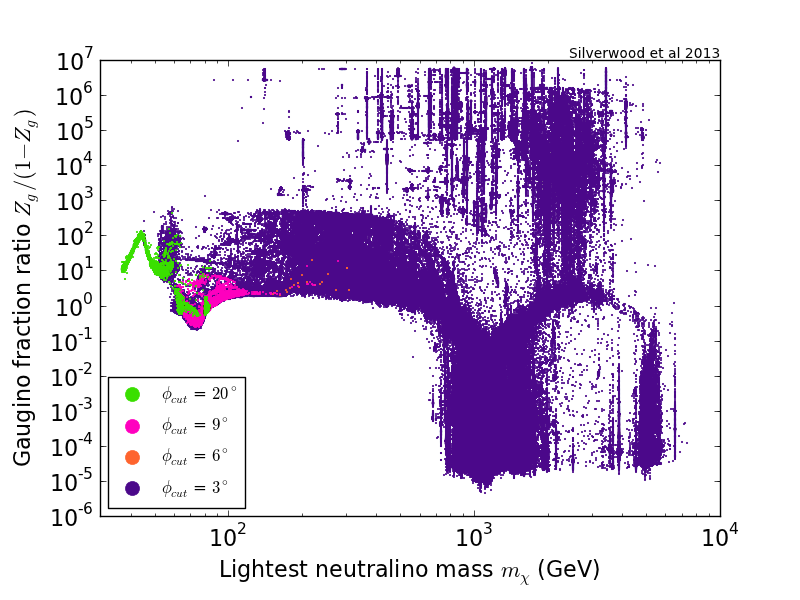}
		\caption{Gaugino fraction ratio $Z_g / (1-Z_g)$ of the lightest neutralino against lightest neutralino mass $m_\chi$,  for points derived from explorations of the MSSM-25 parameter space. Colour coding indicates the optimal $\phi_\text{cut}$ angle found for each point.}
		\label{Article_Phi_cuts_2_Gaugino_mx1}
	\end{minipage}
	\hspace{0.5cm}
	\begin{minipage}[t]{0.5\linewidth}
		\centering
		\includegraphics[width=\textwidth, trim = 0 10 40 0, clip=true]{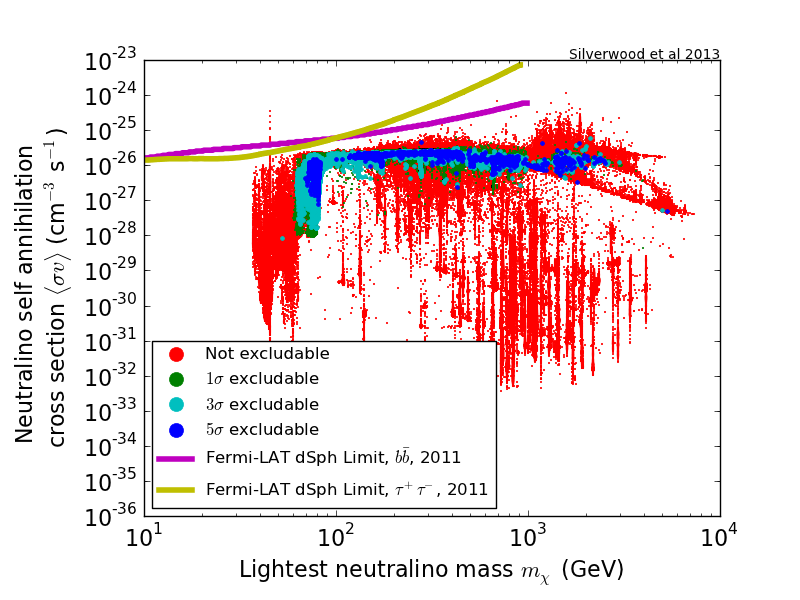}
		\caption{Neutralino self annihilation cross section $\langle\sigma v\rangle$ against lightest neutralino mass $m_\chi$,  for points derived from explorations of the MSSM-25 parameter space. Colour coding indicates predicted IceCube-86 model exclusion CL. 95\% CL upper limits placed on the neutralino self annihilation cross section $\langle\sigma v\rangle$ by measurements from Fermi-LAT assuming annihilation into either $b\bar{b}$ or $\tau^+ \tau^-$ final states are displayed as magenta and yellow lines respectively \citep{FermiLAT2011}.}
		\label{NMvAchRates1_sigma_v}
	\end{minipage}
\end{figure}

Of the 6.02 million points we analysed, 5.75 million had an optimal $\phi_\text{cut} = 3^{\circ}$, 32 had $\phi_\text{cut} = 6^{\circ}$, 156,493 had $\phi_\text{cut} = 9^{\circ}$, and 117,713 had $\phi_\text{cut} = 20^{\circ}$. Figure \ref{Article_Phi_cuts_2_Gaugino_mx1} shows the gaugino fraction ratio against the lightest neutralino mass, with colour coding indicating the optimal $\phi_\text{cut}$ found for each of our 6.02 million points. An optimal $\phi_\text{cut}$ of $3^\circ$ is dominant at high neutralino mass $m_\chi$, while larger optimal $\phi_\text{cut}$ values become increasingly prevalent at lower neutralino masses. This confirms our earlier statement in Subsection \ref{SubsectionLikelihoodCalculation}: higher-mass neutralinos tend to produce higher energy neutrinos, which are more accurately reconstructed by IceCube and therefore more densely clustered around the solar position. In this case, a smaller cut angle reduces the number of background events more than it reduces the number of signal events. Conversely, lower-mass neutralinos will tend to produce lower energy neutrinos and so have a wider angular dispersion of reconstructed events; a wider cut angle is then required to maximise the number of signal events included in the analysis. 

Figure \ref{NMvAchRates1_sigma_v} shows the self annihilation cross section for the lightest neutralino $\langle \sigma v \rangle$ against its mass $m_\chi$. Also shown are 95\% CL limits on the neutralino self annihilation cross section derived from Fermi-LAT observations of dwarf spheroidal satellite galaxies (dSphs) of the Milky Way, assuming annihilation into either $b\bar{b}$ or $\tau^+ \tau^-$ final states \citep{FermiLAT2011}.  The vast majority of points lie below these exclusion lines, including all areas of blue and cyan, showing that IceCube-86 has exclusion capability at 99\% CL or better beyond that of recent Fermi-LAT measurements. 

\begin{figure}
	\centering
	\includegraphics[width=\linewidth, trim = 60 40 95 0, clip=true]{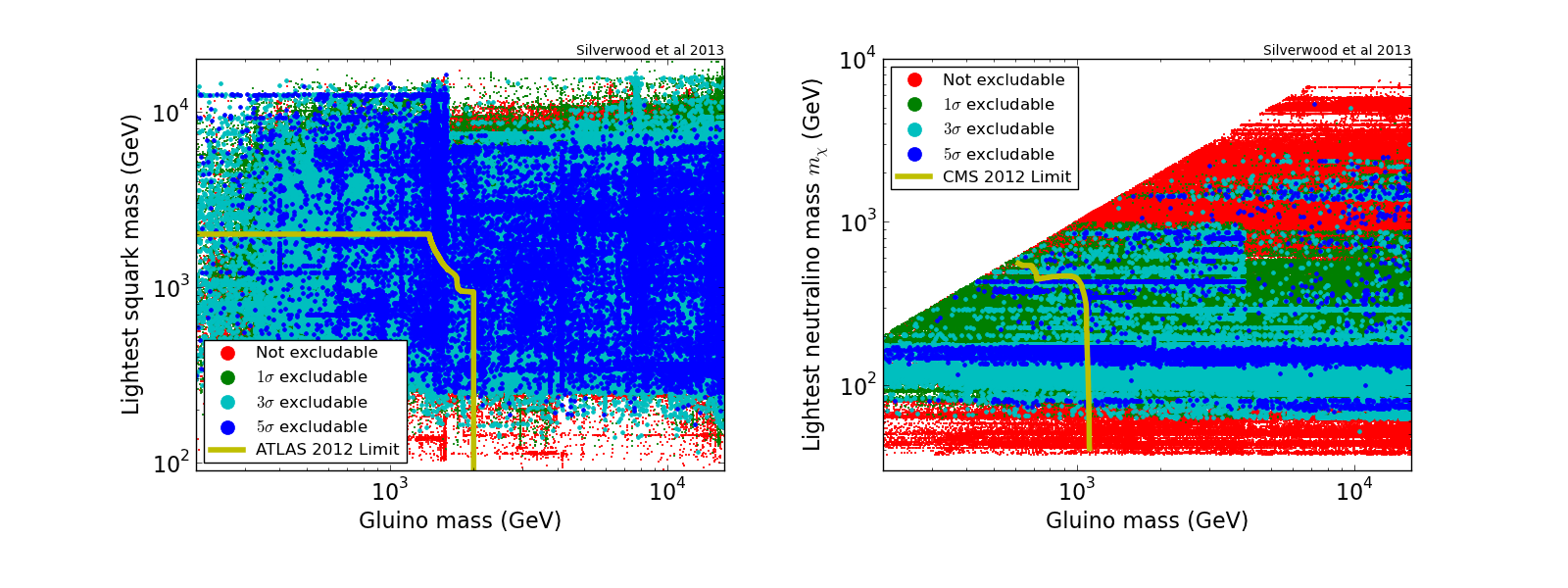}
	\caption{Lightest squark mass (left) and lightest neutralino mass (right) against gluino mass for points derived from explorations of the MSSM-25 parameter space. In the left panel the points to the bottom left of the magenta line are excluded by ATLAS at 95\% CL, based on searches for direct production of coloured sparticles and their decay into jets and missing transverse energy (MET).  The limits are derived by assuming that all sparticles are heavy ($m=5$\,TeV) except for the gluino and first- and second-generation squarks (the masses of which are scanned over), and the lightest neutralino (which is taken as approximately massless) \citep{ATLAS2012}.  Note that our MSSM-25 models do not rely on this assumption. In the right panel, points to the bottom left of the magenta line are excluded by CMS at 95\% CL. This line is the combination of the best simplified model spectrum limits from the `razor' \citep{CMS_PAS_SUS-11-024} and $M_{\text{T2}}$ \citep{CMS_SUSY_MT2_2012} analyses on gluino pair production and decay to diquarks $+$ MET.  These simplified analyses assume that each gluino in the pair decays identically directly to $q \bar{q} \tilde{\chi}^0$, i.e.\ that there are no intermediate states causing cascade decays of gluinos to neutralinos.  This implies that squarks, charginos and the other neutralinos are either heavier than the gluino, or almost degenerate in mass with the lightest neutralino. Note that our MSSM-25 models do not rely on any such assumptions. Colour coding indicates predicted IceCube-86 model exclusion CL.}
	\label{ArticlePlots1_GluinoMass_v_LightestSquarkMass}
\end{figure}

Figure \ref{ArticlePlots1_GluinoMass_v_LightestSquarkMass} Left shows the lightest squark mass against the gluino mass for the points found during the MSSM-25 parameter space exploration. Also shown is the 95\% CL exclusion limit derived from 4.71\,fb$^{-1}$ of data at $\sqrt{s} = 7$\,TeV from the ATLAS experiment at the Large Hadron Collider (LHC) \citep{ATLAS2012}. This limit assumes a simplified set of SUSY models where all sparticles are given masses of 5\,TeV except for the gluino, first- and second-generation squarks, and the lightest neutralino. The first- and second-generation squark masses are scanned over, and the lightest neutralino is taken to be approximately massless. The ATLAS analysis also assumes that the coloured sparticles are produced directly and have direct decays into jets and missing transverse energy. Points below and to the left of the ATLAS limit curve are excluded under these assumptions. The MSSM-25 models we have produced are not based upon any such assumptions. The large number of blue and cyan points above and to the right of this limit curve show that IceCube-86 is sensitive at better than 99\% CL to many models beyond the 95\% CL exclusion capability of ATLAS.

Figure \ref{ArticlePlots1_GluinoMass_v_LightestSquarkMass} Right shows the lightest neutralino mass against the gluino mass for our database of MSSM-25 points. Also shown is the 95\% CL exclusion limit derived from 4.7\,fb$^{-1}$ of data taken at $\sqrt{s} = 7$\,TeV by the CMS experiment at the LHC, with points below this line considered excluded \citep{CMS_PAS_SUS-11-024, CMS_SUSY_MT2_2012}.  The line is the combination of the simplified model spectrum limits from the `razor' \citep{CMS_PAS_SUS-11-024} and $M_{\text{T2}}$ \citep{CMS_SUSY_MT2_2012} analyses on gluino pair production and decay to diquarks $+$ missing transverse energy (MET).  The former is based on limits from 4 $b$-tagged jets $+$ jets $+$ MET shown in Figure 6 Top Left of \citep{CMS_PAS_SUS-11-024}, and the latter from the $\tilde{g} \rightarrow b \bar{b} \tilde{\chi}^0$ channel using the $M_{\text{T2}}b$ analysis, shown in Figure 6 Top Right of \citep{CMS_SUSY_MT2_2012}.  Such simplified model spectra assume that the gluinos in the pair decay identically, and that their decays to neutralinos proceed directly via $\tilde{g}\rightarrow q \bar{q} \tilde{\chi}^0$ only, without any cascade decays via intermediate states.  This implies that other sparticles are either heavier than gluinos, uncoloured, or almost degenerate in mass with the LSP (such that their decay to neutralinos is kinematically suppressed or forbidden).  In practice this makes such limits applicable to the MSSM-25 only in cases where each squark, neutralino and chargino is either heavier than the gluino, or very close to the LSP mass; our MSSM-25 models do not make this assumption, so the CMS limit must be interpreted with quite some care.  Nonetheless, the large number of blue and cyan points above and to the right of this limit curve suggest that IceCube-86 has the capability to exclude many models at better than 99\% CL that are beyond the current 95\% CL reach of CMS.

\begin{figure}[tbp]
	\begin{minipage}[b]{0.5\linewidth}
		\centering
		\includegraphics[width=\textwidth, trim = 20 10 40 0, clip=true]{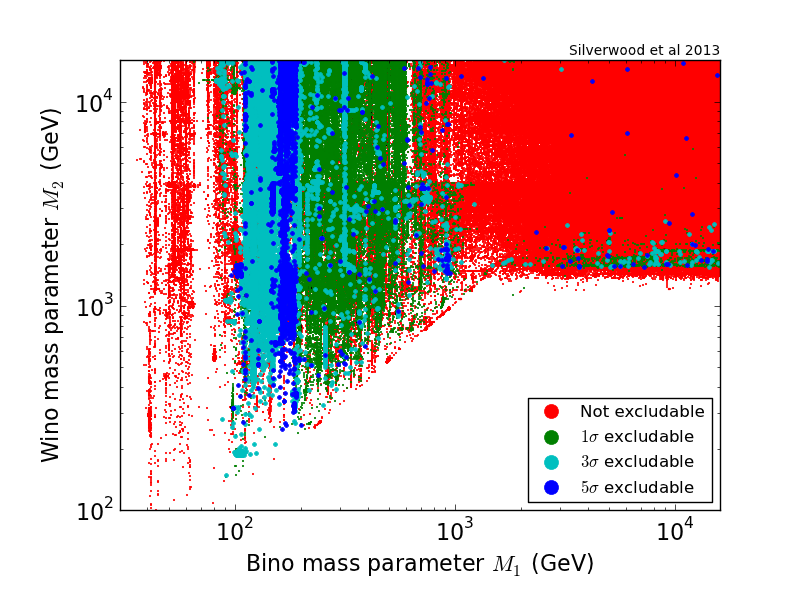}
		\caption{Wino mass parameter $M_2$ against bino mass parameter $M_1$ for points derived from explorations of the MSSM-25 parameter space. Colour coding indicates predicted IceCube-86 model exclusion CL. This plot is similar to the plot generated by the $M_3$ against $M_1$ parameter combination.}
		\label{PvP1_Outm1_v_Outm2}
	\end{minipage}
	\hspace{0.5cm}
	\begin{minipage}[b]{0.5\linewidth}
		\centering
		\includegraphics[width=\textwidth, trim = 20 10 40 0, clip=true]{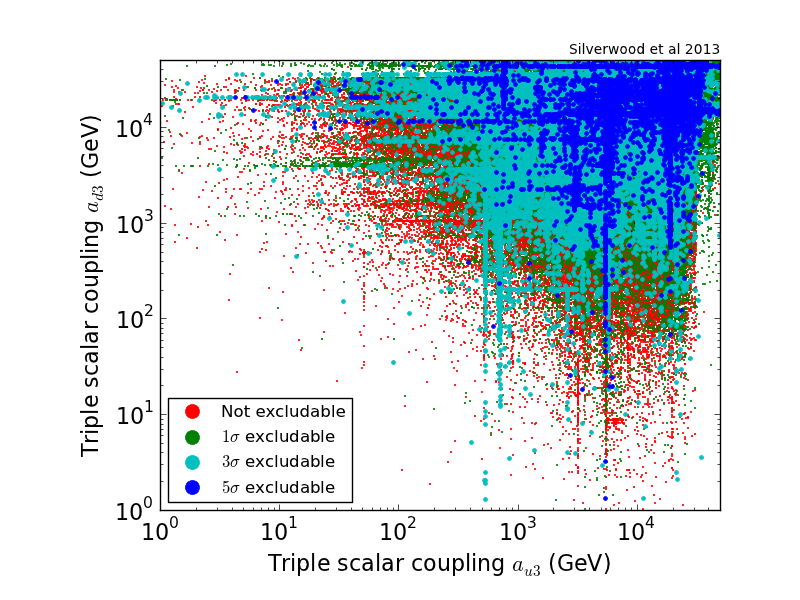}
		\caption{Triple Scalar Couplings $a_{d3}$ against $a_{u3}$ for points derived from explorations of the MSSM-25 parameter space. Colour coding indicates predicted IceCube-86 model exclusion CL. This plot is a representative sample of plots generated by other combinations of triple scalar couplings.}
		\label{PvP15_Outatm_v_Outabm}
	\end{minipage}
\end{figure}

\begin{figure}[tbp]
	\centering
	\includegraphics[width=\linewidth, trim = 60 30 95 0, clip=true]{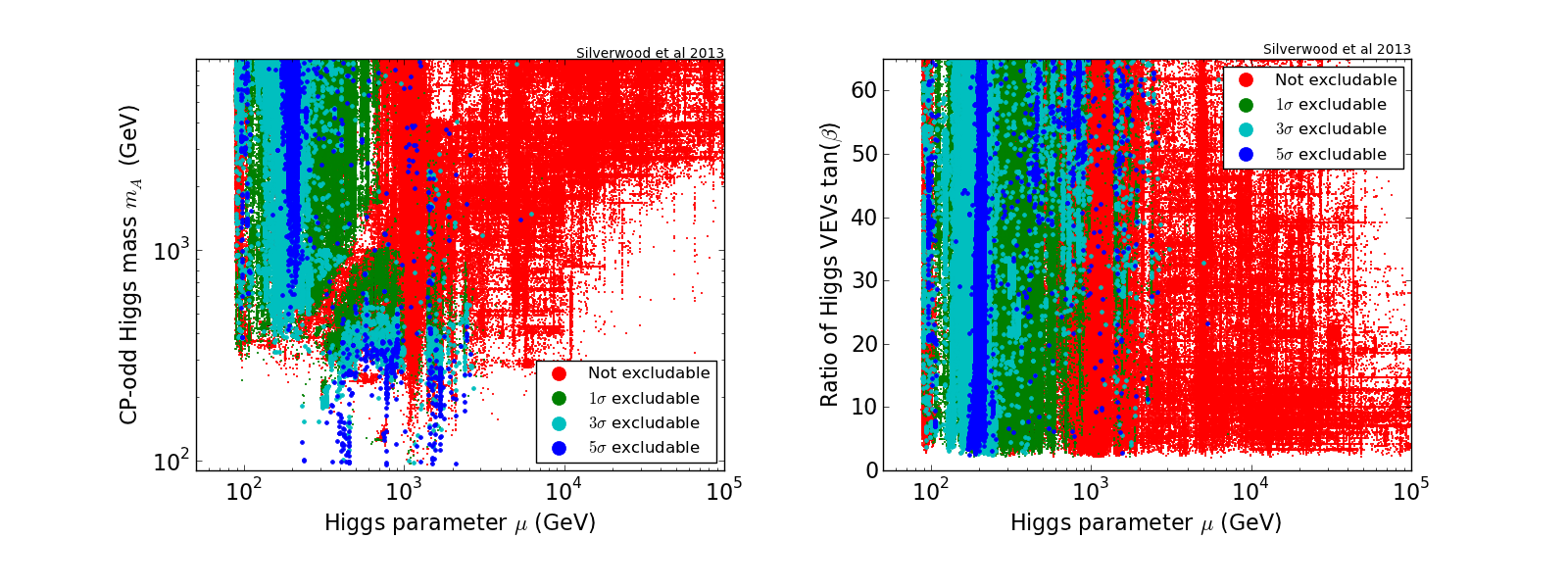}
	\caption{CP-odd Higgs mass $m_A$ (left) and ratio of Higgs VEVs $\tan{\beta}$ (right) against Higgs parameter $\mu$ for points derived from explorations of the MSSM-25 parameter space. Colour coding indicates predicted IceCube-86 model exclusion CL.}
	\label{PvP_A2_Higgs_mu_v_mA_tanbe}
\end{figure}
	
\begin{figure}[tbp]
	\centering
	\includegraphics[width=\linewidth, trim = 60 30 95 0, clip=true]{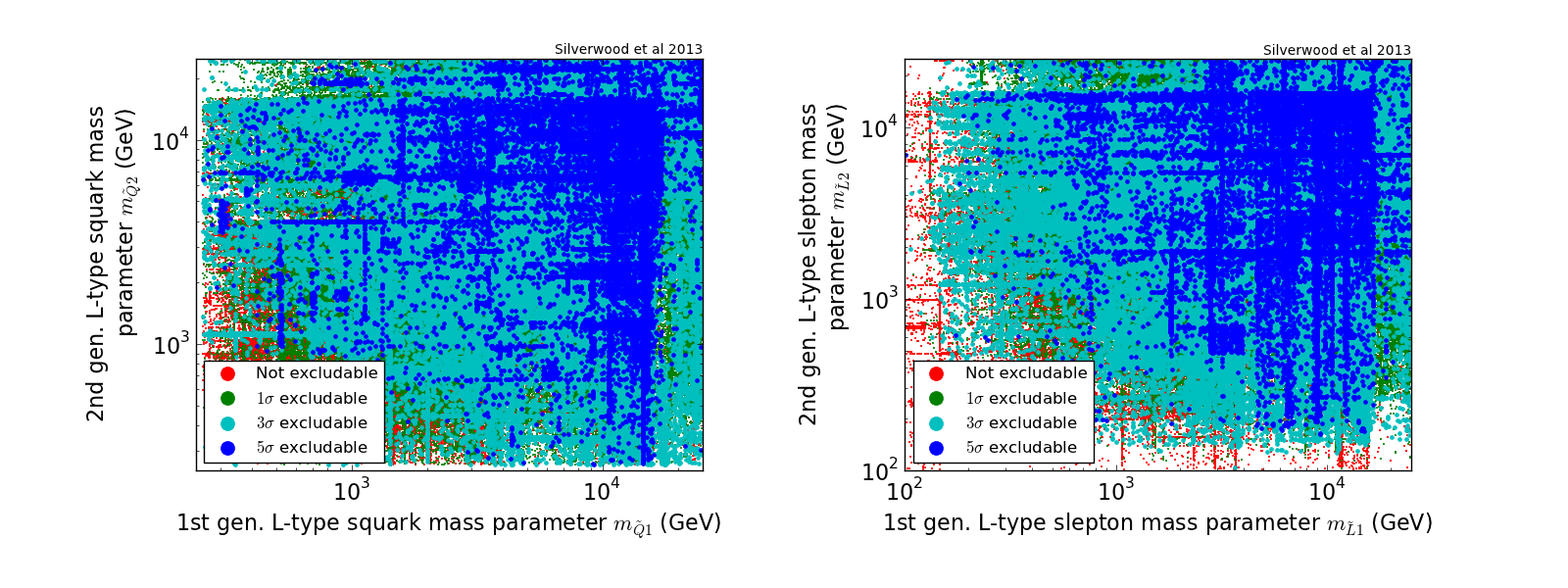}
	\caption{Second generation L-type squark mass parameters $M_{\tilde{Q}_2}$ against first generation L-type squark mass parameter $M_{\tilde{Q}_1}$ (left), and second generation L-type slepton mass parameter $M_{\tilde{L}_2}$ against first generation L-type slepton mass parameter $M_{\tilde{L}_1}$ (right) for points derived from explorations of the MSSM-25 parameter space. Colour coding indicates predicted IceCube-86 model exclusion CL. These plots are a representative sample of plots generated by other combinations of squark and slepton mass parameters respectively.}
	\label{PvP_A3_Sfermion_Mass_Parameters}
\end{figure}

Figures \ref{PvP1_Outm1_v_Outm2} to \ref{PvP_A3_Sfermion_Mass_Parameters} show the exclusion CL across a selection of MSSM-25 parameters. In cases where combinations of parameters from the same category give similar plots, for instance squark mass parameters $M_{\tilde{Q}_1}$ against $M_{\tilde{Q}_2}$ and $M_{\tilde{Q}_1}$ against $M_{\tilde{Q}_3}$, we take one such plot as a representative sample. Several parameters display a strong influence on exclusion CL, as witnessed by concentrations of blue and cyan points. These concentrations can be found where 100\,GeV $\lesssim M_1 \lesssim$ 200\,GeV, (Figure \ref{PvP1_Outm1_v_Outm2}), around $\mu = 200$\,GeV (Figure \ref{PvP_A2_Higgs_mu_v_mA_tanbe} Left and Right), and in the region bounded roughly by 100\,GeV $\lesssim  m_A \lesssim$ 500\,GeV and 400\,GeV $\lesssim \mu \lesssim1200$\,GeV (Figure \ref{PvP_A2_Higgs_mu_v_mA_tanbe} Left). There is a clustering of high exclusion CL points at high values of the triple scalar couplings, as seen in Figure \ref{PvP15_Outatm_v_Outabm}. The sfermion mass parameters (Figure \ref{PvP_A3_Sfermion_Mass_Parameters})  and $\tan{\beta}$ (Figure \ref{PvP_A2_Higgs_mu_v_mA_tanbe} Right) show no influence over exclusion CL, as can be seen from the roughly even distribution of blue and cyan points over their entire ranges.

Here it is useful to recall our earlier note at the start of Section \ref{SectionResultsDiscussion} regarding the interpretation of these plots. Points of high exclusion CL are plotted over those of lower exclusion CL, i.e.\ below the clusters of blue and cyan points there will be red and green points. Some parameter values produce large numbers of models that can be excluded at high CL by IceCube-86, but one cannot claim that IceCube-86 can exclude \textit{all} points with those parameter values at high CL. The only correct way to make a statement about which parts of a parameter space are preferred or excluded, and at what CL, is to perform a full statistical global fit, as described in \citep{IC22methods}. 

The distribution of points across the parameter space shows some distinct edges at certain parameter values: at $M_2 = 4000$\,GeV in Figure \ref{PvP1_Outm1_v_Outm2}, at $m_A = 4000$\,GeV in the left panel of Figure \ref{PvP_A2_Higgs_mu_v_mA_tanbe}, and the `box within a box' features seen in Figures \ref{PvP15_Outatm_v_Outabm} and \ref{PvP_A3_Sfermion_Mass_Parameters}. These are artefacts of the scanning method and the parameter limits for various scans. For instance, we performed scans with limits on $M_2$ of $\pm 4000$\,GeV and $\pm 16000$\,GeV, and so there are sharp cut-offs at these values. Similarly, we performed scans with limits on $m_A$ of $\pm 4000$\,GeV and $\pm 8000$\,GeV, leading to similar features in Figure \ref{PvP_A2_Higgs_mu_v_mA_tanbe} Left. We performed a large number of scans with sfermion mass parameters limited to $\pm 16000$\,GeV, but only a few with limits expanded to $\pm 25000$\,GeV. Thus the region bounded by this first limit was much more heavily sampled, and so has a higher density of points. Similarly the majority of scans we performed had limits on triple scalar couplings $a_{\text{u}3}$, $a_{\text{d}3}$, $a_{\text{e}1/2}$,  and $a_{\text{e}3}$ set to $\pm 32000$\,GeV, but only a few had these limits expanded to $\pm 50000$\,GeV. 

\begin{figure}[tbp]
	\centering
	\includegraphics[width=0.5\textwidth, trim = 20 10 40 0, clip=true]{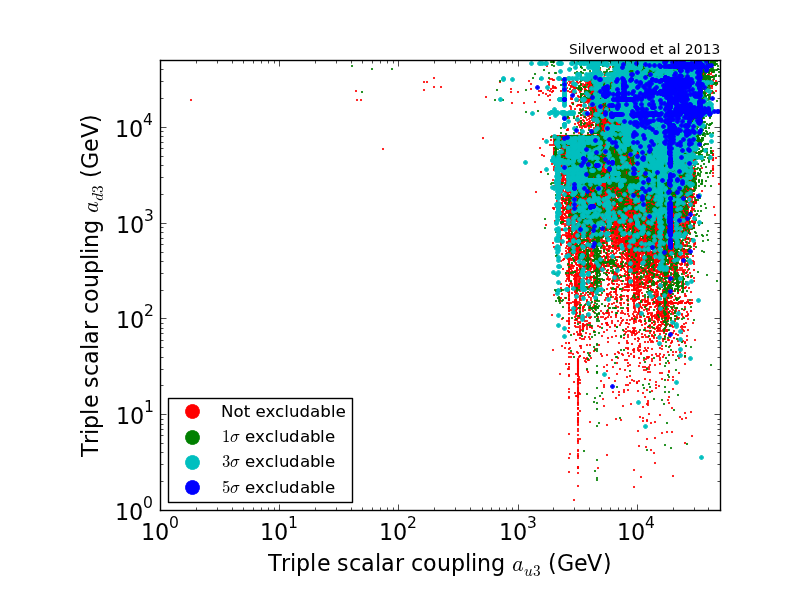}
	\caption{Triple Scalar Couplings $a_{d3}$ against $a_{u3}$ for points derived from explorations of the MSSM-25 parameter space and with CP-even Higgs mass in the range 124.6 - 126.8\,GeV as indicated by recent ATLAS and CMS measurements \citep{ATLAS_Higgs_2012, CMS_Higgs_2012}. Colour coding indicates predicted IceCube-86 model exclusion CL. Comparison to Figure \ref{PvP15_Outatm_v_Outabm} reveals the impact of the CP-even Higgs mass cut. This plot is a representative sample of plots generated by other combinations of triple scalar couplings with CP-even Higgs mass cuts applied.}
	\label{PvP15_Outatm_v_Outabm_Higgs_Cut}
	\hspace{0.5cm}
\end{figure}

Recent results from the ATLAS and CMS experiments indicate the existence of a particle compatible with the Standard Model Higgs boson and supersymmetric CP-even Higgs bosons within mass ranges of 125.2 - 126.8\,GeV and 124.6 - 126.2\,GeV respectively \citep{ATLAS_Higgs_2012, CMS_Higgs_2012}. To gauge the impact of this measurement upon our results we performed a cut on our database to extract all parameter space points producing a mass for at least one of the CP-even Higgs boson $H^0$ or $h^0$ within the combined ATLAS-CMS range of 124.6 - 126.8\,GeV. Of the 6.02 million parameter space points in our database 1.22 million survived this cut. This cut does not alter the overall distribution of points across parameter space, except in the case of the triple scalar coupling $a_{u3}$. Comparing the pre-cut distribution shown in Figure \ref{PvP15_Outatm_v_Outabm} to the post-cut distribution as shown in Figure \ref{PvP15_Outatm_v_Outabm_Higgs_Cut} we can clearly see a large number of points with $a_{u3}$ less than 2000\,GeV are eliminated. 

\section{Conclusions}
Our explorations of the MSSM-25 parameter space have yielded 6.02 million parameter space points that produce neutralino LSPs, satisfy constraints imposed by accelerator data, and produce a neutralino relic density within the $2\sigma$ bounds of the 7-year WMAP measurement. Calculations of model exclusion CL for these points show that the 86-string configuration of IceCube will be able to rule out or detect at $3\sigma$ (99.7\% CL) or better a significant number of models that are beyond the 90\% CL exclusion limits of current direct detection experiments such as SIMPLE, COUPP, and XENON100, and beyond the current 95\% CL exclusion limits of Fermi-LAT, ATLAS, and CMS.  

\begin{acknowledgments}
P.S. is supported by the Lorne Trottier Chair in Astrophysics, an Institute for Particle Physics Theory Fellowship and a Banting Fellowship, administered by the Natural Science and Engineering Research Council of Canada. C.S. is supported by the Swedish Research Council (VR) through the Oskar Klein Centre and the Department of Physics \& Astronomy at the University of Utah. J.E. and K.H. are supported by VR through respective grants 621-2010-3301 and 621-2010-3705, and via the Oskar Klein Centre. J.A.A. and A.M.B. acknowledge support from the Marsden Fund Council from New Zealand Government funding, administered by the Royal Society of New Zealand.

\end{acknowledgments}

\bibliographystyle{JHEP_pat}
\bibliography{MSSM25_Scans}

\end{document}